# Implementation of Live Reverse Debugging in LLDB

Anthony Savidis[1,2] (Corresponding Author), Vangelis Tsiatsianas[2]
[1] *Institute of Computer Science, FORTH, GR*
[2] *Computer Science Department, University of Crete, GR*
{as@ics.forth.gr, vangelists@csd.uoc.gr}

SUMMARY

Debugging is an essential process with a large share of the development effort, being a relentless quest for offensive code through tracing, inspection and iterative running sessions. Probably every developer has been in a situation with a clear wish to rewind time just for a while, only to retry some actions alternatively, instead of restarting the entire session. Well, the genie to fulfill such a wish is known as a reverse debugger. Their inherent technical complexity makes them very hard to implement, while the imposed execution overhead turns them to less preferable for adoption. There are only a few available, most being off-line tools, working on recorded, previously run, sessions. We consider live reverse debuggers both challenging and promising, since they can fit into existing forward debuggers, and we developed the first live reverse debugger on top of LLDB, discussing in detail our implementation approach.

## 1. INTRODUCTION

Debugging is the systematic process of detecting and fixing bugs within computer programs and can be summarized by two main steps: bug detection and bug fixing, as outlined under Figure 1. In this context, the bug detection process can be very demanding, requiring numerous repeated execution rounds, while it heavily relies in the deployment of debugging instruments to trace, examine, and analyze the program behavior. Currently debuggers play an important role, since their capabilities and features may affect how easily bugs are eventually captured in the source program.

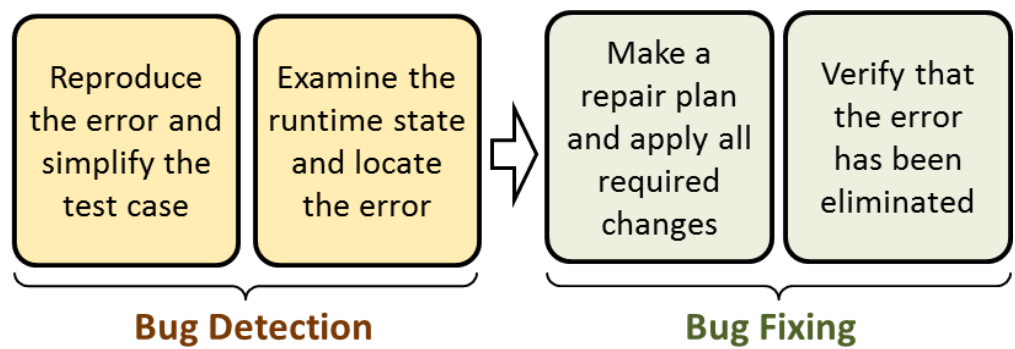

Figure 1. An outline of the overall debugging process (adapted form [1]).

The general debugger architecture, capturing the primary abstractions and leaving out the lower-level details, is illustrated under Figure 2, involving primarily three key components: (i) the debugger backend, being usually the most language or platform dependent part of the story; (ii) the debugger frontend, which tends in most situations to suit and adapt to the backend functions; and (c) the debugger user-interface, which deploys the frontend to offer interactive debugging functionality.

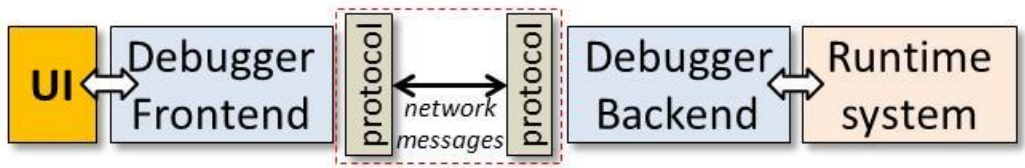

Figure 2. High-level architecture of a debugger system.

The low-level linkage between the frontend and the backend is not standard and differs across language platforms, and may involve network communication, including the necessary protocol parsers at each end, or may be very system-specific, as with the case of LLDB [29] and typical C++ applications we address in this work.

*1.1 Forward Debugging*

Forward debuggers execute the target program instructions in the same way they would normally run without the debugger being attached. Thus, instructions execute in sequence, unless they concern branches, function returns, or exception handling, where the control-flow is accordingly updated. Then, functionality is provided to control execution, such as to set breakpoints, pause, or continue, together with some variations of stepwise tracing. Finally, extra features not related to execution control are offered for state inspection (watch points, expression evaluation and call stack).

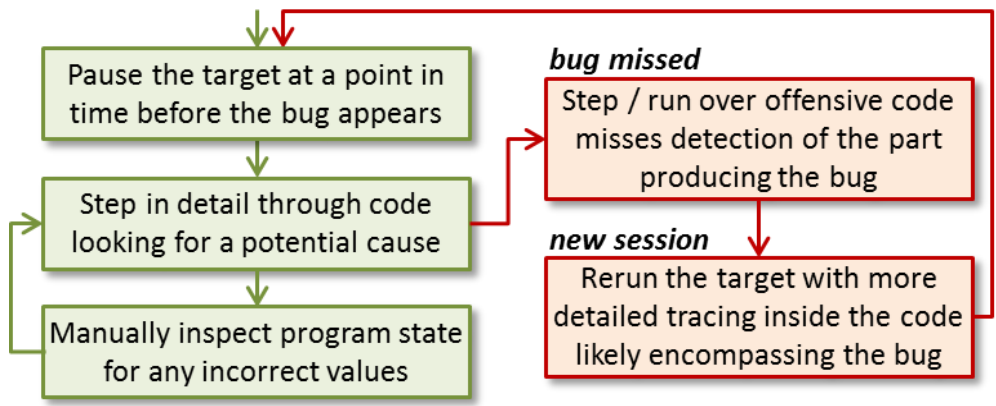

Figure 3. Program tracing approach for bug detection in a typical forward debugging process.

*1.2 Reverse Debugging*

Reverse debugging refers to the ability of debuggers to somehow step backwards we well, by precisely restoring the state of the program to that earlier point in time. Such debuggers may work in a live manner, that is while the target program is executed, or in an offline one, based on a previously captured trace of the target's execution, including appropriate timestamped state recordings [4].

*1.2.1 Offline Reverse Debugging*

Effectively, off-line tools are very powerful, interactive, *state analysis instruments*, enabling to inspect in detail the entire execution history of a runtime session, just like a movie that can be played forward and backward on demand, however, without ever altering the plot and the flow. They are known as time-travel debugging tools [9], [11] and the majority of existing reverse debugging tools are off-line. Due to the way offline reverse debuggers actually function they are known as record-and-replay tools. The target is recorded once, and then, after the execution finishes, the reverse debugger acts on the recording, instead of a live target (see Figure 4).

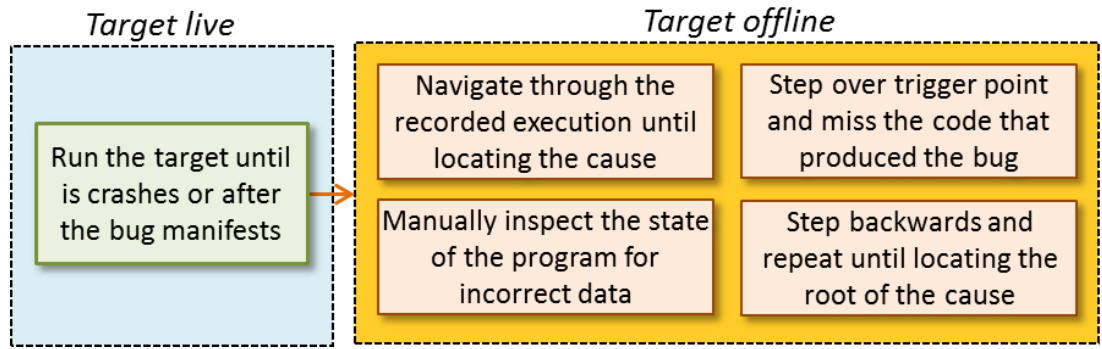

Figure 4. Typical record-and-replay offline debugging process, with error reproduction performed on a live target, while state examination and replay on an offline one.

The benefits of this approach lie on the fact that the incorrect behavior needs to be reproduced only once and afterwards the user can freely examine the recording at their own pace and time. In addition, the recorded execution trace may also be transferred to another computer for the analysis process, a feature of great importance for debugging issues in production systems, especially when the access to the machine where the fault appeared is not possible or allowed. For instance, a customer that encounters an issue with a software system may collect such a trace and send it to the software vendor for further examination. A direct drawback of this approach, however, is that the trace may contain sensitive data, e.g., passwords, security keys or other company secrets that should remain locally in the customer computer.

### 1.2.2 Live Reverse Debugging

Essentially, in such debuggers, stepping back should behave as an *undo* function, applying in reverse sequence the opposite effect of a series of earlier executed instructions. Then, users may retry forward execution of any undone program part and take an alternative control flow, not bound back and forth to the exact traces (see Figure 5). Thus, live reverse debugging offers *more chances for exploratory testing*.

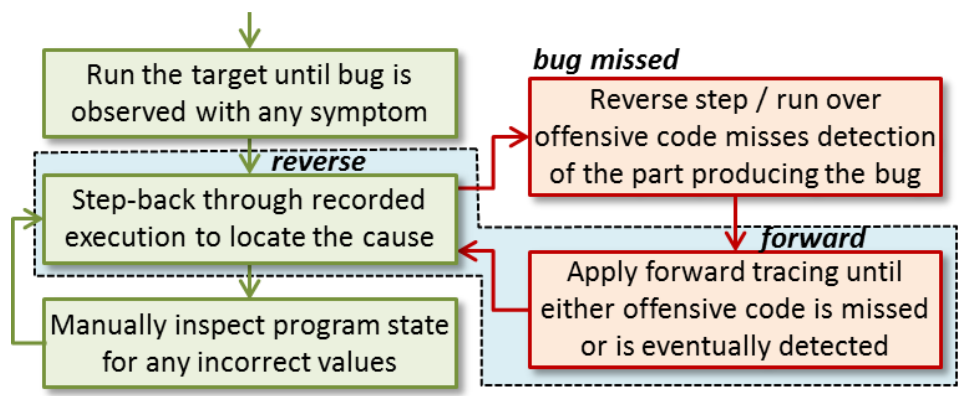

Figure 5. The essence of live reverse debugging is the dynamic blending and interplay of reverse and forward program tracing processes (indicated in the dotted shaded area).

Effectively, live reverse debugging is important because: (a) state modifications after undoing can alter the previous control-flow, enabling testing alternative execution paths back-and-forth, until the offensive code is located, even when missed during the initial forward session; (b) manual updates in the offended state may be combined with iterative backward and forward sessions, to live experiment with potential state resolutions when the defects are essentially incorrect assignments (code updates, however, always require rebuild and rerun sessions). Another potential use-case for such debuggers is to enable users originally unfamiliar with a particular codebase or library to browse through the code during execution, via successive forward and backwards stepping actions, combined with manual state updates, in order to examine and better comprehend a particular piece of code or gain direct insight on how the components of a software system interact with and affect each other.

The flexibility of live reverse debugging is shown with a very small example under Figure 6, where an alternative execution scenario is caused and live tested, after reverse stepping, by purposefully modifying specific variables.

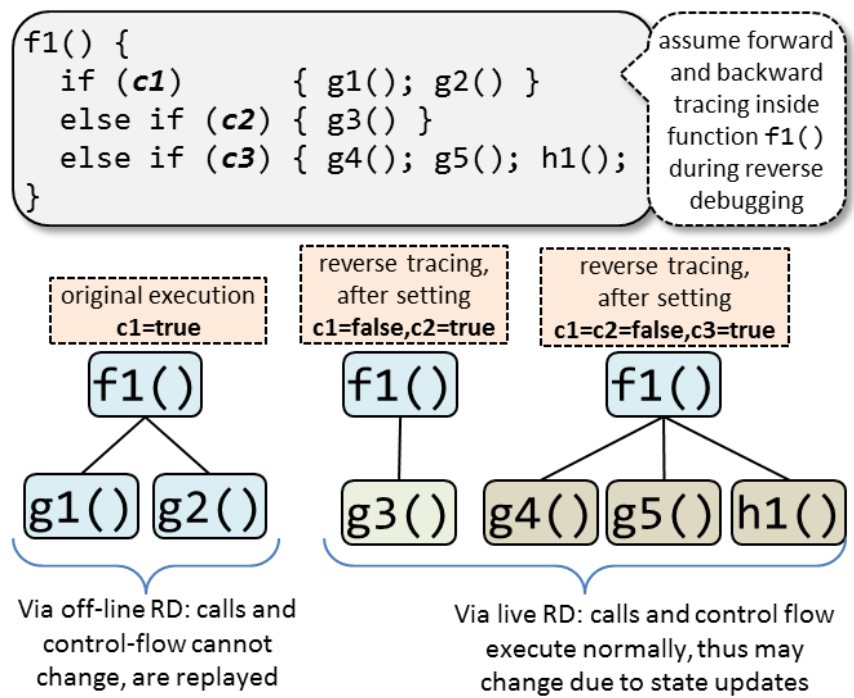

Figure 6. In live reverse-debugging (LRD) the control-flow (thus function invocations too) may change, when compared to record-and-replay reverse-debugging.

Overall, live reverse debugging is very challenging, however, bringing directly on the table questions regarding the ability to reverse every type of execution event. The clear answer to this is *not in general*. Indeed, there are numerous cases where it is either technically infeasible, or proved impractical, or even sometimes undefined, for various categories of execution events. For instance, operations on system resources, like output devices or the network, cannot be reversed. Also, if we manage to restore their state to an earlier image, all perceivable effects, once observed or tracked by independent software systems or humans, are theoretically set as non-reversible (we cannot yet undo human memory writes, and once we intervene on the behavior of third-party systems we are known as viruses).

In summary, we consider live reverse debugging not a panacea for all debugging scenarios, but a very handy and targeted tool, with clear limitations, but also with well-defined advantages, mostly linked to the ability of exploratory testing, back and forth in the control flow, via alternative execution paths, all within a single debugging session. Our work concerns live reverse debugging, and is to our knowledge, the first systematic and comprehensive implementation effort on top of the LLDB API.

*1.3 Contribution*

Our primary contribution is to elaborate our implementation practice and experience in building the first live reverse debugger in LLDB, mainly using the public client-APIs, while also introducing a few required intrinsic updates that we discuss in detail, but no other lower-level modifications.

All live reverse functionality supported relies thoroughly on the standard approach of LLDB to inject debugger logic with the documented hooks, and we did not deviate from that. Along these lines, we discuss the details of our implementation, and most importantly we explain the general technical approach in hand-crafting a live reverse debugger from scratch, while justifying our key design choices.

The repository[1] of our work with full implementation is publicly available, and also a brief video[2] demonstration which shows the LRD working in real practice.

*1.4 About the LLVM Debugger*

LLDB [1] is a debugger for C, Objective-C and C++ programs, although derivative works have added support for other programming languages, notably for Swift LLVM [2] and Rust [3]. LLDB is written in C++, like most of the LLVM Project, and is maintained as a subproject within the latter.

As noted in its homepage, *"...LLDB is built as a set of reusable components which highly leverage existing libraries in the larger LLVM project, such as the Clang expression parser and LLVM disassembler"*.

The LLDB debugger is currently used in a variety of settings and is the default debugger both in the Apple Xcode IDE and the Android Studio IDE. LLDB uses the Clang compiler frontend [5] to evaluate user expressions written in the same source language as the running target program.

2. REQUIREMENTS
*2.1 Capturing Program Execution*

In order to step backwards, the reverse debugger needs to trace the execution of the target program and capture its full or partial state, along with any side-effects from non-deterministic input sources, such as results of certain system calls, interrupts, input device events or network requests and changes made to shared process memory. There are several approaches to saving the state of the target, differentiated by whether they modify the target executable and by the granularity of the trace.

---

[1] https://github.com/vangelists/llvm-project

[2] https://vimeo.com/419351406

### 2.1.1 Binary Instrumentation

Binary instrumentation (see Figure 7) refers to the injection of extra code in the original program for the purpose of usually inserting additional behavior or monitoring its performance. In the context of reverse debuggers, the instrumentation code is injected near control flow instructions and those modifying the program state. This way, they are able to capture the execution order of the target's machine or byte code instructions, along with any modifications made by the latter to variables, registers and memory. Using this method implies either recompiling the source code of the software or executing the target using Just-In-Time (JIT) techniques.

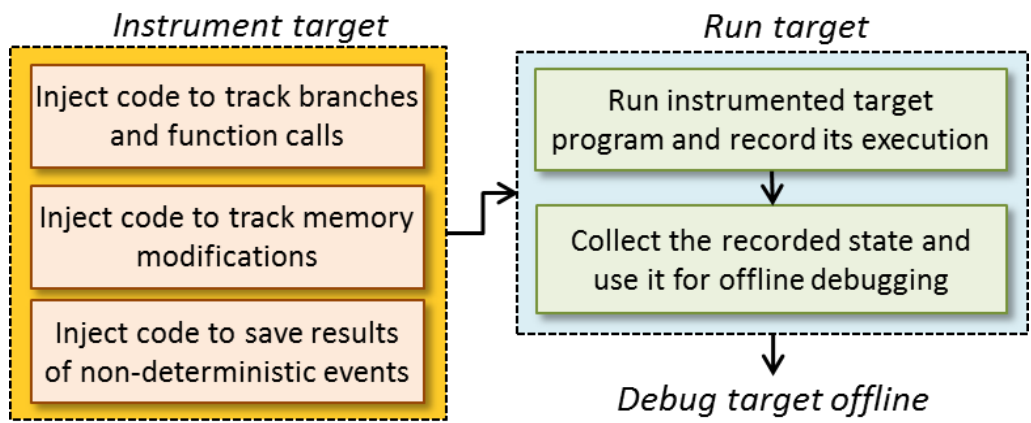

Figure 7. Binary instrumentation for the purposes of an offline reverse-debugging process.

In the first case, the source code has to be available and the programmer should recompile the software, something considerably increasing the size of the executable, which may not be acceptable in some cases, as for example in embedded systems. In the second case, recompilation is optional and some reverse debuggers have successfully applied a JIT-based approach. An example of a commercial reverse debugger using JIT to trace the execution of the target program is the former UndoDB [6], now named UDB [7] that is provided by the "*undo*" company. On the other side, binary instrumentation provides the benefit that there is no need for a particular system or debugger to be in place, during the execution of the program that would be responsible for tracing the target. This is important for production systems, where attaching a debugger to catch the bug may be impossible, but a trace can be captured and saved for future offline analysis.

### 2.1.2 Time-Based Snapshots

Another approach is to run the original target program unmodified and periodically take snapshots of its state. This approach allows for both fine-grained and coarse-grained tracing, and highlights the tradeoff between the accuracy of the trace, in terms of making sure to capture every state change, versus its performance, memory and storage overhead. Capturing a snapshot of the target at longer intervals allows for faster execution and longer traces at the expense of higher risk to fail recording the root cause of the bug, be it an incorrect assignment or an invalid control flow path, as we can see in the example of Figure 8. On the other hand, shorter snapshot intervals translate to greater overhead, but also a greater chance to capture all modifications of interest in the program, along with the cause of the bug.

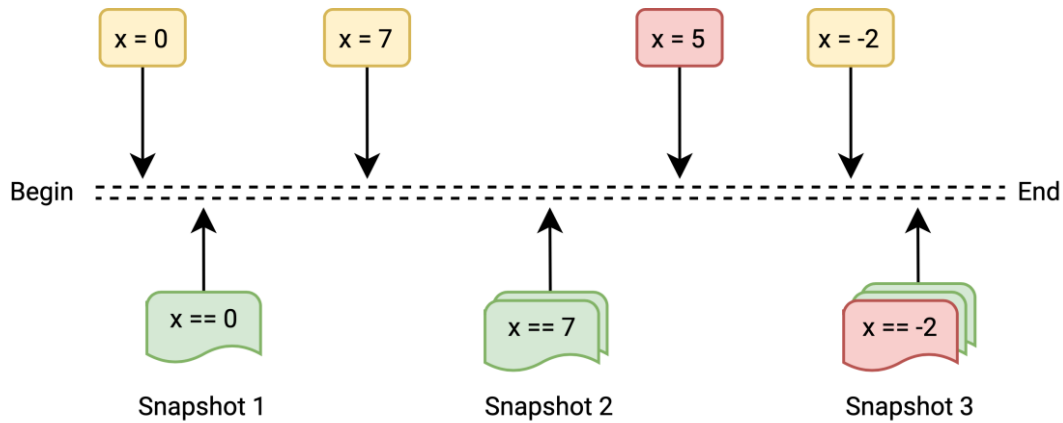

Figure 8. An example of time-based snapshots (upper part shows actual writes), missing assignment to a variable (*x=5*) due to choosing a relatively-large snapshot interval.

### 2.1.3 Event-Based Snapshots

As already discussed in the previous section, time-based tracing has the increased risk of omitting the state changes of interest, leading to failures in detecting the reason a bug actually occurs. Event-based tracing concerns software or hardware systems that capture every state change of the target program, by recording its state before and after each machine code instruction is executed, tracking memory accesses and any sources of non-determinism. Thus, in this context, an *event* is considered the execution of an instruction, a memory access, an interrupt, a system call or any other action that results in the modification of the target program variables, registers and memory, or a control flow change. Revisiting the example of Figure 8, we observe that event-based snapshots are more suitable for cases when accuracy matters more than the potential performance or memory overhead, as shown under Figure 9.

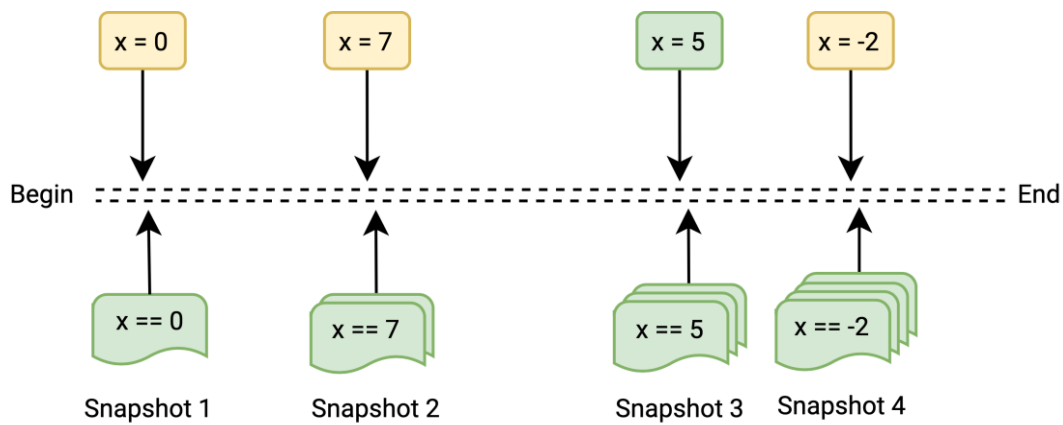

Figure 9. Capturing precisely every single assignment through event-based snapshots.

### 2.2 Full vs Partial Snapshots

In taking snapshots, either the state of a running target may be captured as a whole, or could be restricted to capturing only the side-effects of non-deterministic stimuli. Snapshotting the entire state of the target at a certain execution point may allow the debugger to restore that full state without executing the target at all, however, it also translates to far greater memory and storage overhead. Conversely, partial snapshots incur smaller or minimal memory and storage overhead, yet the debugger is required to rerun the target, up to the exact runtime point of interest, in order to reconstruct the original state. This is achieved by committing the effects of deterministic instructions, while replacing all non-deterministic instructions, and also all calls to non-deterministic functions or system calls, with their respective results, exactly as they have been captured during the traced execution of the target.

*2.3 Limitations of Reverse Debuggers*

There are certain limitations regarding reverse debuggers, concerning not only the problems debugger users face, but also technical issues faced by their developers.

From the point of view of debugger developers, a major obstacle is an inherent technical difficulty in their implementation, due to the many low-level factors affecting state recording during target execution, like shared memory, multiple threads and child processes, storage and network accesses and cross-platform support. In addition, reverse debuggers have to keep up with frequent updates to operating system kernels, since side-effects of all system calls must be known and handled.

On the end-user side, it is the severe performance and memory overhead on the execution of the running target that hinders their overall adoption. Also, all known solutions seem to forcefully serialize multi-threaded target programs in order to record their execution, thus slowing down such targets even further, by an extra factor equal to the number of the originally parallel threads. As a result, even though there are various free and commercial solutions, reverse debuggers are not enthusiastically received, compared to traditional forward debuggers, and we may even argue they still remain unknown to the vast majority of programmers.

## 3. ARCHITECTURE

We use *single-step mode*, a special mode in CPUs that forces the processor to execute one instruction at a time and then stop with an interrupt. This is also known as single-step interrupt, named *TYPE 1* interrupt, and is associated to the single-step execution when debugging a target program. This mode is almost exclusively used by debuggers to implement forward-stepping at the instruction level, and we use it in our live reverse debugger to trace the execution of the target program. Enabling this mode is generally achieved by setting a processor flag, known as the *trap flag*.

*3.1 Debuggee Representation*

In this section, we briefly introduce how the debuggee is represented in the basic LLDB software architecture, at the code level, since we rely on it, before proceeding to elaborate on the details of our implementation.

A target program consists of one or more processes, which may have a number of threads (see Figure 10, left part). The root class representing the target program is `Target`, with `Process` and `Thread` naturally representing running processes and threads. Besides execution information for the debuggee, the `Target` also stores metadata for all shared and dynamically loaded libraries using `ModuleList`, which holds a collection of `Module` instances. Each `Thread` (see Figure 10, right part) owns a different version of CPU registers, represented in LLDB by `RegisterContext`, with an instance owned by `Thread`. The API of `RegisterContext` makes use of `RegisterValue` to export the register values to other classes. Additionally, `Thread` holds an instance of `Unwind`, which is in charge of visiting and unwinding the entire call-stack when the thread stops and populating the `StackFrameList`, which represents the call-stack holding an instance of `StackFrame` per actual frame.

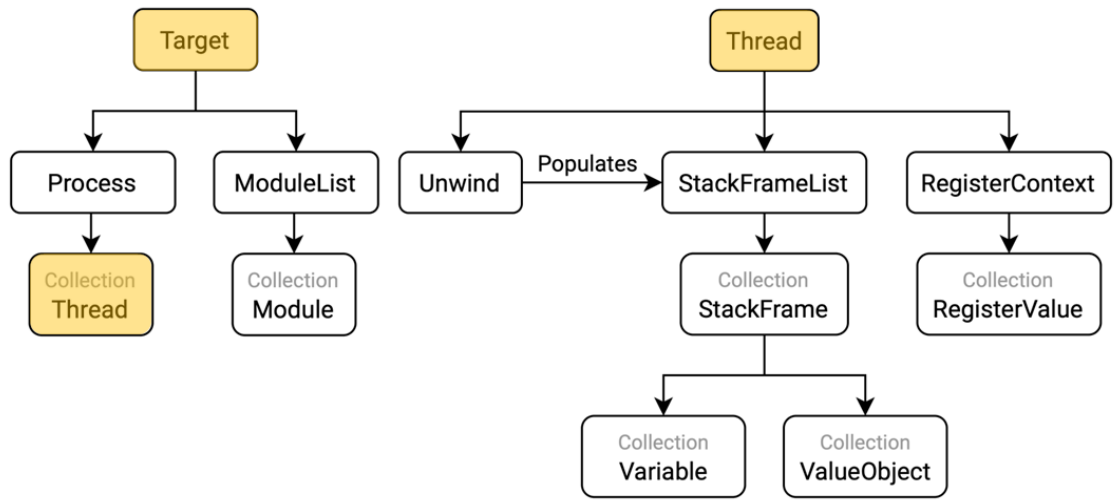

Figure 10. The primary *Target* class owning a collection of *Thread* instances and the way frame data are kept within *Thread* instances.

The `StackFrame` holds information about all variables belonging to a respective call frame. In particular, `StackFrame` maintains a list of `Variable` instances, encompassing symbolic information for each variable (e.g., name, type, scope and source language). The values of variables are kept in a separate list with `ValueObject` instances inside `StackFrame`.

*3.2 Thread Plans*

The debugging runtime behavior of each thread (step into function, step out of function, run until address, continue, etc.) is controlled by behavior atoms represented by `ThreadPlan`. Each thread maintains a separate stack of thread plans that is populated based on user commands or internal decisions before execution resumes.

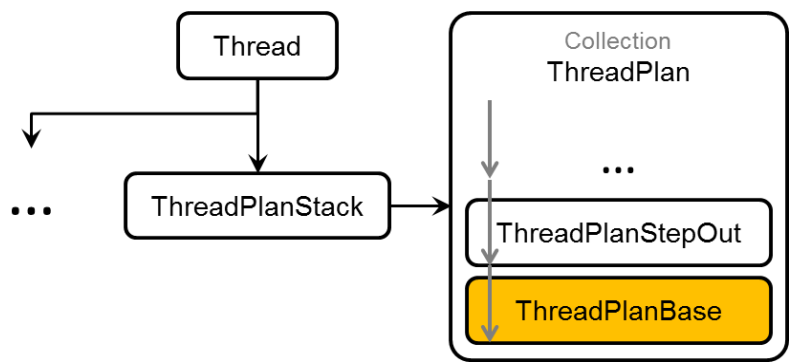

Figure 11. The *ThreadPlanStack* containing a (stack) of *ThreadPlan* instances, with a *ThreadPlanBase* object always reserved at the bottom - three dots represent the rest of *Thread* class members.

When the thread stops, all plans are queried, starting from the top of the stack, asking if their *mission has been accomplished*. If this is true, these thread plans are marked as completed and are later discarded. A thread plan can also be discarded before its mission is accomplished, in case anything goes wrong during its execution, such as when a function invocation leads to a system crash. The thread plan stack of any `Thread`, by design, can never be empty; there is always a plan at the bottom of the stack, known as the *base plan*, being always an instance of `ThreadPlanBase`. The base plan is responsible to take care of generic stop events, such as hitting a breakpoint, raising signal or an exception.

*3.3 Thread Plan Tracers*

A rather obscure feature of the LLDB implementation is the ability to force a thread enter single-step mode, something made possible via the `ThreadPlanTracer` and also by respective derivatives. More specifically, an instance of `ThreadPlanTracer` is held by the *resident base plan* instance (always at bottom of stack) and is queried *after every stop-point*, checking whether the thread is currently in single-step mode. Besides deciding the execution mode of the current thread, the tracer also acts to *log* its results directly on the output console.

Originally, the `ThreadPlanTracer` class has been designed with the purpose to serve as a debugging utility for LLDB developers, by displaying the disassembly of the current stack frame in every stop-point. However, the ability of a thread plan tracer to enforce single-step execution mode on the running thread, along with the ability to run arbitrary code *right before* each machine code instruction is executed, enabled us to use it as the basis for all reverse debugging functionality incorporated within LLDB, as part of the reported work. The rest of the discussion concerns our implementation on top of this infrastructure.

*3.4 Trace Points and Bookmarks*

In our implementation, the class representing a point in time within the live recorded history is `Tracepoint`, used by our `ThreadPlanInstructionTracer`, being subclass of the original LLDB `ThreadPlanTracer`, which holds `Tracepoint` instances, essentially our information carriers for execution snapshots. In order to aid navigating through the recorded target execution, we allow reverse debugger users to mark tracepoints using *bookmarks*, which can be optionally named (i.e. tagged).

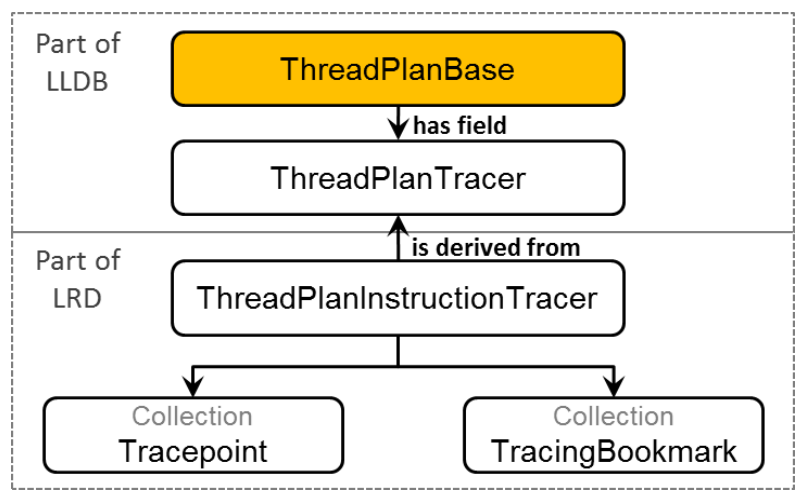

Figure 12. The *ThreadPlanInstructionTracer, Tracepoint* and *TracingBookmark* classes, part of our LRD (live reverse debugger) implementation on top of LLDB.

Bookmarks are stored in a collection of `TracingBoomark` instances and provide the ability to easily jump between tracepoints of interest, enabling a faster and more efficient debugging process. The sample console output of Figure 13 shows the creation of a bookmark and the listing of all available bookmarks within the current (live) recording. If the user is currently at a tracepoint marked by a bookmark, then the enlisted bookmarked is prefixed with an asterisk as a navigation hint. As shown, the bookmark user functions begin with `bm` (shortcut, fully qualified prefix is actually `thread tracing bookmark`) followed by the specific command name, and are used in command-line to `create` a bookmark with name (`-n` option) and then to enumerate all available bookmarks with `list` command.

```
(lldb) bm create -n "Initialization of 'x'"

Created bookmark at tracepoint 1: "Initialization of 'x'"

(lldb) bm list

Current bookmarks:
* 1: "Initialization of 'x'"
   └─ Tracepoint 276: memory_access`main + 313
      at memory_access.cpp:51:5, address = 0x0000000100001809
  2: "Incrementing 'x' by the value of 'global'"
   └─ Tracepoint 326: memory_access`main + 558
      at memory_access.cpp:65:12, address = 0x00000001000018fe
```

Figure 13. An LRD debug session with a couple trace bookmarks being added and viewed.

In the code of Figure 14 we show the `Tracepoint` data structure, playing a key role in our reverse debugger since it is the actual information stored by the tracer for every point in time a snapshot must be captured (i.e. a snapshot is a `Tracepoint` instance).

```
struct Tracepoint {
    // The ID of the tracepoint
    Thread::TracepointID   id;
    // The values of stack frame registers
    RegisterValues         registers;
    // The values of stack frame variables
    VariableValues         variables;
    // The contents of a heap region before being
    // overwritten by an instruction that modifies that
    // region, if applicable
    llvm::Optional<HeapData> heap_data;
    // The depth of the deepest stack frame
    uint32_t               frame_depth;
    // The stack frames present when the thread stopped
    StackFrames            frames;
    // The stop reason of the thread
    lldb::StopInfoSP       stop_info;
    // The thread plans completed by the stop
    ThreadPlans            completed_plans;
    // The source line at this point in time, if available
    uint32_t               line;
};
```

Figure 14. The *Tracepoint*, being the LRD runtime snapshot data structure.

## 4. IMPLEMENTATION
### 4.1 Capturing Snapshots

As was briefly mentioned in at the beginning of the previous section, the reverse debugging features added on top of LLDB are based on forcing the thread to always execute in single-step mode and break back into the debugger directly before executing each instruction. The component responsible for recording, managing and restoring the state of the thread is our new `ThreadPlanInstructionTracer` class which replaces the original default `ThreadPlanTracer` tracer of LLDB within `ThreadBasePlan` instance. The latter, as said earlier, is always reserved (resident) at the bottom of the `ThreadPlan` stack. This way, the debugger becomes able to capture a snapshot of the thread at every machine code instruction.

```cpp
void ThreadPlanInstructionTracer::CaptureSnapshot (void) {
   StackFrameList& frame_list = *m_thread.GetStackFrameList();
   // Save register values
   RegisterValues registers;
   DoForEachStackFrame(frame_list, [&](StackFrame &frame) {
      registers.push_back( GetStackFrameRegisterValues(frame) );
   });
   // Save current stack frames
   StackFrames frames = frame_list.CheckpointStackFrameList();
   // Save values of stack frame variables
   VariableValues variables;
   DoForEachStackFrame(frame_list, [&](StackFrame &frame) {
      variables.push_back( GetStackFrameVariableValues(frame) );
   });
   // Save thread state and current source line
   StopInfoSP      stop_info = m_thread.GetStopInfo();
   ThreadPlanStack completed_plans = GetCompletedPlanStack();
   const uint32_t  line = GetSymbolContext().line_entry.line;
   // Append snapshot to history and update tracepoint index
   m_timeline.emplace_back(++m_current_tracepoint, registers,
      variables, frames, stop_info,
      completed_plans, line);
   // Perform any pending heap data backup
   SaveRecentlyStoredHeapDataIfNeeded();
}
```

Figure 15. Implementation of the LRD main snapshot capturing logic.

In the following sections, we briefly discuss the methods followed to capture the total state of the thread at a particular point in time (implementation provided under Figure 15), and also track modifications made by the program and the library functions to the heap, along with the reasoning behind some of these decisions, where required.

#### 4.1.1 Registers, Variables and Stack Frames

The values of all registers and variables for every active stack frame are being backed up at each stop, except for the *exception* state registers, which are always ignored, due to limitations in affecting their values when restoring the state of the thread. Saving the register values is achieved by storing instances of `RegisterValue` alongside the

LLDB-specific *ID* of the register. Respectively, backing up the variables of a stack frame requires saving both the `ValueObject` containing the actual data of the variable and the `Variable` object containing its corresponding metadata. Also, modified registers and variables are marked, enabling the user to list modifications via the `thread tracing modification list` command of the LRD.

Finally, the `StackFrameList`, alongside its contained `StackFrame` objects are being deep-copied, in addition all related metadata, such as the index of the currently selected stack frame.

*4.1.2 Heap Modifications*

Tracking modifications to the heap is a process involving a few steps (see Figure 16). Firstly, disassembling the instruction about to be executed and requesting information regarding its potential behavior, via the architecture-specific disassembler plugin. In case the instruction in question is recognized as one that may *store*, then the destination operand is translated into a virtual memory address, within the address space of the process owning the thread, and the instruction mnemonic is used to extract the number of bytes about to be stored to the destination memory location. Afterwards, the tracer verifies that the destination address corresponds to the heap, i.e., does not belong to the stack or correspond to any known symbol or code.

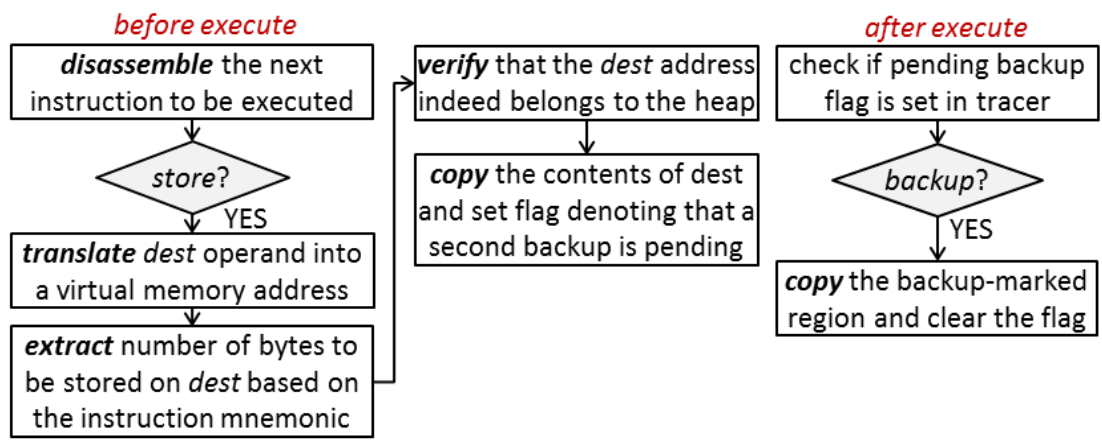

Figure 16. Steps involved in handling heap modification instructions.

Doing so, involves the following two steps: (i) information about the stack, within the process memory, is extracted from the `Process` instance using the current value of the stack pointer to tell the limits of the stack region; and (ii) the `ModuleList` containing the metadata about all loaded images is fetched from the `Target` and traversed, with every `Module` queried to find out whether the destination address resolves to a global variable or a function.

Given that the address corresponds to the heap, all contents of that memory location are saved before the instruction is executed, in order to back up the current (old) contents and a flag is set denoting that this heap region awaits a second backup. Eventually, after the instruction is executed and control returns to the debugger, the tracer recognizes that a backup is pending and saves the current (new) contents of the aforementioned heap region, enabling the debugger to undo and redo the modification. Finally, in a similar manner to registers and variables, modified heap regions are also marked, allowing the user to enlist all modifications to a region of interest using the `thread tracing modification list` command.

*4.1.2.1 Modifications by Library Functions and System Calls*

Calls to external library functions and system calls are be decision not traced, either to speed up execution or simply because it is impossible to do so. Consequently, the tracer will not break before every single internal instruction within the binary code of such calls to track potential state modifications. However, their overall effect must be appropriately tracked.

In order to mitigate this problem, we decided that the tracer keeps an explicit list of system library functions and system calls, registered at start-up, along with the necessary functions that are responsible for handling their side-effects (i.e. the memory backup logic). For instance, our tracer is aware of the system library function `memcpy()`, since we have inserted it in the registry, together with the respective handler function, which we manually implemented to also address the particular semantics and the calling convention of the platform where the target runs. In particular, when calling a function in a Unix-like system on *x86-64*, the first argument is placed in `RDI`, the second in `RSI` and the third in `RDX`. Thus, in this particular case, when a call to `memcpy()` is encountered, the tracer extracts the destination address from `RDI` and the size of the copied data from `RDX` and makes sure to copy the exact heap region about to be overwritten, while also backing up the new contents of that region after the call returns.

Clearly, this is a wrap-around requiring an explicit registration of every heap-affecting library or system call with its respective heap-backup mechanism, but we consider it to be a reasonable choice, since it scales-up and must be implemented once.

*4.1.3  Thread State and Source Location*

After saving all the data related to the state of the target program, the tracer backs up the state of its parent `Thread` instance that must be restored, either when stepping backwards and replaying instructions, or when the execution of the target continues forward. This information includes the source line at the original point in time, the latest stop reason, and the thread plans up to that point. The stop reason needs to be backed up, because it is modified whenever the user steps backwards, or replays one / many recorded instructions, or triggers the evaluation of an expression.

*4.2 Restoring Snapshots*

During a live debugging session we restore the state of both the target program and the LLDB data structures representing that state. The tracer method for this job is implemented in Figure 17. Restoring the state of the `Thread` owning the tracer is straightforward by reassigning all class fields to their previously backed up values.

```
void ThreadPlanInstructionTracer::RestoreSnapshot (
    Thread::TracepointID tracepoint_id
) {
  // Restore heap data
  if (tracepoint_id < m_current_tracepoint)
    UndoHeapWritesUpTo(tracepoint_id);
  else
    RedoHeapWritesUpTo(tracepoint_id);
  // Update current tracepoint index
  m_current_tracepoint = tracepoint_id;
  // Restore stack frames
  Tracepoint &tracepoint = m_timeline[m_current_tracepoint];
  m_thread.SetStackFrameList(*tracepoint.frames);
  // Restore register and variable values for topmost frame
  RestoreStackFrameState(0);
  // Restore thread state
  if (StopInfoSP stop_info = tracepoint.stop_info; stop_info) {
    stop_info->MakeStopInfoValid();
    m_thread.SetStopInfo(stop_info);
  }
  m_thread.SetCompletedPlanStack(tracepoint.completed_plans);
}
```

Figure 17. Implementation of the LRD snapshot restoration logic.

### 4.2.1 Restoring Heap Modifications

Restoring the heap is accomplished by undoing or reapplying the modifications made by each store instruction sequentially, up to the point in time where the user has opted to navigate. If a heap region has since been unmapped, then the restoration of the old contents fails and the user is warned that all history associated with that particular heap region will be discarded, since it is no longer needed (see Figure 18).

```
(lldb) thread tracing step-back
error: Failed to write process memory:
       memory write failed for 0x100125000
error: The heap region 0x100125000 - 0x100125003 is no longer
       accessible, thus all recorded history for this area will
       be discarded.
```

Figure 18. Informing the user about deallocated heap regions (error messages).

On the other side, if the heap region in question is still mapped and thus writable, but its contents have been invalidated or moved, e.g. through a call to `free()` or `realloc()`, respectively, then that particular heap region ends up in an undefined state for which the user remains unaware. Enabling `tracing-jump-over-deallocation-functions` in LLDB settings will cause the debugger to not execute traced calls to known deallocation functions, such as `free()` and `munmap()`, increasing the number of heap regions that will ultimately be available, since such regions will not be invalidated and reclaimed or unmapped.

That is achieved by *detecting calls to such functions and replacing the opcode of the call instruction with a* `nop`, until the thread moves on to the next instruction and the opcode of the call instruction is restored. In order to replace the instruction, the tracer first needs to figure out the number of bytes that form its opcode and then replace those bytes in memory with an equal number of bytes for the `nop` instruction, as is defined by the underlying microarchitecture. Currently, the tracer maintains all possible representations of `nop` for the *x86-64* microarchitecture (see Figure 19).

```
constexpr uint8_t i386_nop_opcodes[][max_i386_nop_opcode_size] {
   { 0x90 },
   { 0x66, 0x90 },
   { 0x0f, 0x1f, 0x00 },
   { 0x0f, 0x1f, 0x40, 0x00 },
   { 0x0f, 0x1f, 0x44, 0x00, 0x00 },
   { 0x66, 0x0f, 0x1f, 0x40, 0x00, 0x00 },
   { 0x0f, 0x1f, 0x80, 0x00, 0x00, 0x00, 0x00 },
   { 0x0f, 0x1f, 0x84, 0x00, 0x00, 0x00, 0x00, 0x00 },
   { 0x66, 0x0f, 0x1f, 0x84, 0x00, 0x00, 0x00, 0x00, 0x00 }
};
```

Figure 19. Encodings of *nop* instruction in the *x86-64* CPU family.

This replacement operation trick is mostly transparent to the user since the `nop` instruction is only visible in the disassembly and only while the thread is stopped at the replaced call instruction. It should be mentioned that identifying reallocated pages is indeed possible via a custom allocator or special tracking of reallocation functions, such as `realloc()`, however, resolving this issue was not included in the scope of this particular project and, as a result, a solution was not considered. *This is just a feature that was not included in our implementation, while in general it can be directly resolved in the context of live reverse debuggers*.

*4.2.2   Restoring Stack Frames, Registers and Variables*

The full call-stack as such is never being backed up, since the actual memory is never saved, in order to minimize the memory footprint of the tracer. Instead, only the `StackFrameList` held by the `Thread` is saved and restored each time.

Then, after the saved `StackFrameList` is restored, the values of all registers and variables need to be also restored. Regarding the stack frame variables, restoring the saved `ValueObject` for each variable is enough in order to *convince* LLDB about the value of a variable and ensure correct resolution during expression evaluation. On the other hand, simply restoring the `RegisterValue` of each register is not sufficient, since LLDB attempts to directly unwind the call stack when the value of a register is to be read or written.

The method we followed to overcome this undesirable behavior was to introduce a new member in `RegisterContext` class, holding the values of *all* saved registers at any given point in time. Then, we bypassed the methods responsible for accessing the registers in `RegisterContext` class by making sure that any requested register value *is always fetched from the tracer* in case the state of the stack-frame owning that `RegisterContext` is actually emulated (i.e. is restored) to mimic a previous point in runtime (i.e. naturally following a reverse debug command by the user).

Finally, when stepping backwards or forward within the recorded execution history, only values of registers and variables accessible by the topmost stack-frame are restored. This method was picked to avoid an unnecessary overhead, since the user is most probably not going to examine all stack frames on every step, especially if there are many of them also belonging to third-party libraries, as is the case with the system loader libraries. As a result, the values for a stack frame other than the topmost one are restored only when that frame is explicitly selected by the user during debugging, realizing a *typical restore on demand* policy.

*4.3 Avoiding Unwanted Symbols*

Executing the target in single-step mode imposes a *slowdown* of at least *1000x* to the target program. This metric by itself looks very surprising, in a negative way, but to be fair, one should also consider that even typical forward debugging may introduce a *100x* deceleration in typical cases. However, this remains a considerable *10x* average slowdown over an already huge overhead due to normal debug-mode execution.

In this context, tracing redundant symbols, such as system library functions, incurs a great memory overhead, since the snapshots take up a considerable amount of time and memory. As a result, our tracer takes a few measures to avoid tracing symbols that need not be traced, in order to speed up execution and minimize the memory footprint. Currently, such symbols are those that belong either to system libraries installed under `/usr/lib/` or to the `std` C++ namespace. Building on this feature, extra function sets could be installed from a catalogue in the LRD, although we did not had time to include this facility as well. However, the user may define a set of additional functions to ignore via the following LLDB settings:

```
target.process.thread.tracing-avoid-symbols-regex
target.process.thread.tracing-avoid-libraries
```

In order to avoid such symbols (functions), the following actions are taken (see also Figure 20). Single-stepping (trap mode) and tracing are suspended exactly before the relevant call instruction, involving a designated (unwanted) function, is executed, and an artificial breakpoint, which is automatically deleted on the first hit, is set at the instruction right after the call. When the call finishes and the breakpoint is reached, the callback of the breakpoint, which resumes single-stepping and tracing, is executed and the breakpoint is directly deleted, allowing the thread to continue running.

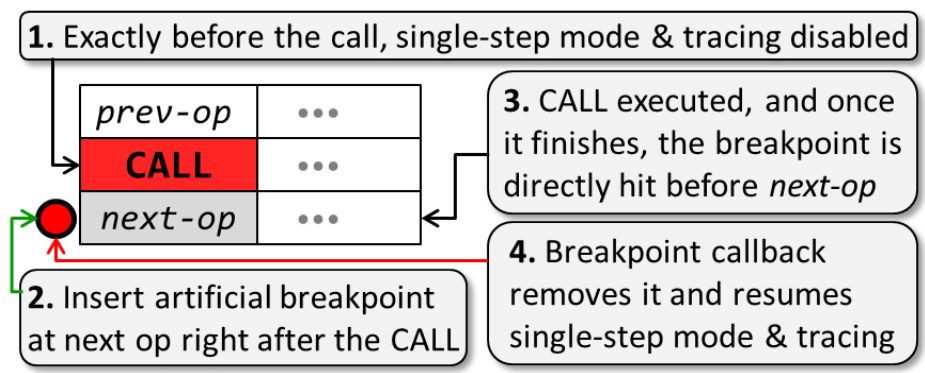

Figure 20. Actions to avoid tracing inside calls to functions marked as *trace-unwanted*.

*4.4 Evaluating User Expressions*

LLDB evaluates user expressions using Clang, the C/C++ compiler that also developed under the umbrella of the LLVM project. More specifically, the user expression is parsed, translated to the source language of the target, compiled into native code for the target platform and then directly injected into the target's memory. Then, the target process is temporarily resumed in order to execute the injected code and calculate the result, which is fetched and displayed to the user in the LLDB console. When the evaluation finishes, the target's memory is cleaned up and the target is restored to the state before the expression evaluation.

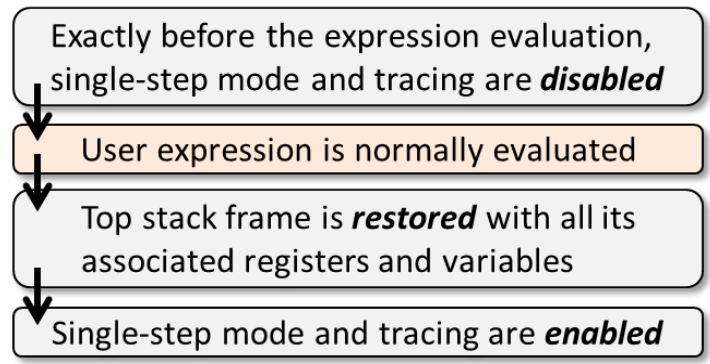

Figure 21. Extra actions added before and after the evaluation of user expressions.

Clearly, polluting the timeline containing the recorded execution history of the target program with snapshots caused due to expressions the user evaluated during the debugging session would be a mistake. Thus, in a similar manner to unwanted symbols that are ignored, tracing and single-stepping are suspended before a user expression is evaluated and resumed right after the evaluation completes (see Figure 21). An important difference to the handling of avoided symbols is that the state of the topmost stack frame is also restored after the evaluation finishes, along with its associated registers and variables, so as to undo any possible modifications that were caused due to the expression evaluation.

*4.5 Respecting User Breakpoints*

When continuing backwards, such as via the reverse versions of the typical continue or run-to commands, the thread steps back until either the beginning of the history is reached or an enabled user breakpoint is actually met. Respectively, when replaying forward, the thread replays recorded instructions until either end of recorded history is reached or an enabled breakpoint is also hit. This is accomplished by firstly traversing the recorded history, before actually restoring the thread to a previous or replayed point in time, and checking explicitly whether the saved program counter in each snapshot corresponds to any currently enabled breakpoint. If a breakpoint is found, thread restoration occurs up-to that exact time spot.

*4.6 Some Caching*

Results of frequent and expensive computations must be cached, so as to improve tracing performance and consequently reduce the slowdown that is imposed on the target. Currently, this translates to caching whether an address corresponds to the heap or the stack and whether a symbol belongs to a library installed under `/usr/lib/`.

*4.7 Reverse Debugging API*

The functionality provided by `ThreadPlanInstructionTracer` is exported internally to the rest of LLDB via `Thread` class. `Thread` acts as the glue between the tracer, that implements the functionality, and the command interpreter, where the user commands are defined. In addition, as also mentioned earlier, the extra API added to `Thread` is also used by the `RegisterContext` class to check whether the stack frame state is currently being emulated so as to fetch saved register values directly from the snapshot captured at a certain point in time. Next we provide an outline of all exported key sub-APIs per category regarding the reverse functionality.

```
class Thread {
      virtual Status StepBack (std::size_t num_statements);
      virtual Status StepBackInstruction (std::size_t num_insts);
      virtual Status StepBackUntilAddress (lldb::addr_t address);
      virtual Status StepBackUntilLine (uint32_t line);
      virtual Status StepBackUntilOutOfFunction (void);
      virtual Status StepBackUntilStart (void);
      virtual Status ReverseContinue (Stream &breakpoint_id);

      virtual Status Replay (std::size_t num_statements);
      virtual Status ReplayInstruction (std::size_t num_insts);
      virtual Status ReplayUntilAddress (lldb::addr_t address);
      virtual Status ReplayUntilLine (uint32_t line);
      virtual Status ReplayUntilOutOfFunction (void);
      virtual Status ReplayUntilEnd (void);
      virtual Status ReplayContinue (Stream &breakpoint_id);

      Status        JumpToTracepoint (TracepointID destination);
      TracepointID  GetCurrentTracepointID (void);
};
```

Figure 22. Reverse execution sub-API.

```
enum class TracedWriteTiming {
      Past, Future, Any
};
class Thread {
      Status ListRegisterWriteLocations (
            Stream& stream, llvm::StringRef register_name,
            std::size_t num_locations,
            lldb::TracedWriteTiming write_timing
      );
      Status ListVariableWriteLocations (
            Stream& stream, llvm::StringRef variable_name,
            std::size_t num_locations,
            lldb::TracedWriteTiming write_timing
      );
      Status ListHeapAddressWriteLocations (
            Stream& stream, lldb::addr_t heap_address,
            std::size_t num_locations,
            lldb::TracedWriteTiming write_timing
      );
};
```

Figure 23. Modification examination sub-API.

```
class Thread {
   llvm::Expected<TracingBookmarkID>
          CreateTracingBookmark (
                 TracepointID    tracepoint_id,
                 llvm::StringRef name = {}
          );
   Status DeleteTracingBookmark (TracingBookmarkID boookmark_id);
   llvm::Expected<const TracingBookmark&>
          GetTracingBookmark (TracingBookmarkID boookmark_id);

   llvm::Expected<const TracingBookmark&>
          GetTracingBookmarkAtTracepoint (
                 TracepointID tracepoint_id
          );
   const TracingBookmarkList
          GetAllTracingBookmarks (void);
   Status JumpToTracingBookmark (TracingBookmarkID boookmark_id);
   Status RenameTracingBookmark (
                 TracingBookmarkID boookmark_id,
                 llvm::StringRef   name
          );
   Status MoveTracingBookmark (
                 TracingBookmarkID boookmark_id,
                 TracepointID      new_tracepoint_id
          );
};
```

Figure 24. Tracepoint bookmarks' sub-API.

## 5. EXAMPLES

We discuss a couple of very basic running scenarios that are simple but demonstrate both the operation as well as the usefulness of the LRD. A more elaborate case study is shown in our video[3] demonstration.

### 5.1.1 Reverting Stack Corruption

In this scenario we demonstrate the ability of a reverse debugger (either offline or live) to operate on a target after the latter has corrupted its call stack, contrary to a traditional forward-only debugger. We are going to use a minimal program, which corrupts its call stack by overwriting its memory via a call to `memset()`, originally aiming to reset the contents of a stack-allocated array, but erroneously using a size greater than that of the array as an argument to `memset()`, as shown under Figure 25. This is an example of a typical buffer overrun error.

---

[3] https://vimeo.com/419351406

```
static void foo() {
      int b[1];
      /* over-run b and corrupt stack */
      memset(b, 0, 20);
      return;
}
int main() {
      foo();
      return 0;
}
```

Figure 25. Sample program that causes stack corruption.

### 5.1.1.1 Via Forward Debugging

Debugging this program with a forward-only debugger would be a very difficult if not impossible task since the debugger is not able to walk and unwind the corrupted stack, thus neither a back-trace nor the variables are available, as shown under Figure 26.

```
* thread #1, queue = 'com.apple.main-thread', stop reason =
EXC_BAD_ACCESS (code=1, address=0x0)
    frame #0: 0x0000000000000000
error: memory read failed for 0x0

(lldb) p b

error: <user expression 1>:1:1: use of undeclared identifier 'b'
```

Figure 26. Output of forward-only debugger after stack corruption occurs.

### 5.1.1.2 Via Reverse Debugging

Our live reverse debugger relies on the recorded snapshots. Thus is capable of fully reconstructing the back-trace even after such a stack infection. This directly allows the developer to normally inspect all stack frame variables after stepping backwards. The output of our LRD as actually used to back-trace via `bt` LLDB command, after stepping back two statements using the LRD command `thread step-back -c 2` (shortcut `ps -c 2` used in this example), exactly after the crash occurred, is provided under Figure 27. After back-trace, we can normally check and print the contents of the `b` array (LLDB command `print p`, or shortcut `p b` used in this example) to only observe it is now back to the initial zeroed state.

```
Process 66392 launched: 'stack' (x86_64)
Process 66392 stopped

* thread #1, queue = 'com.apple.main-thread', stop reason =
EXC_BAD_ACCESS (code=1, address=0x0)
    frame #0: 0x0000000000000000
error: memory read failed for 0x0

(lldb) ps -c 2

* thread #1, queue = 'com.apple.main-thread', stop reason = step
back 2 statements
    frame #0: 0x0000000100000f86 stack`foo() at stack.cpp:5:5
   1    #include <cstring>
   2
   3    static void foo() {
   4        int b[1];
-> 5        memset(b, 0, 20);
   6        return;
   7    }
   8
   9    int main() {
   10       foo();
   11       return 0;

(lldb) bt

* thread #1, queue = 'com.apple.main-thread', stop reason = step
back 2 statements
  * frame #0: 0x0000000100000f86 stack`foo() at stack.cpp:5:5
    frame #1: 0x0000000100000f64 stack`main at stack.cpp:10:5
    frame #2: 0x00007fff6e04ccc9 libdyld.dylib`start + 1

(lldb) p b

(int [1]) $0 = ([0] = 0)
```

Figure 27. Output of reverse debugger after stack corruption.

```
int main() {
     bool return_zero = false;
     if (return_zero) {
          return 1;
     } else {
          return 0;
     }
}
```

Figure 28. Sample program for basic control error.

### 5.1.2 Modifying Control Flow

We use another very simple example with a branch statement and a control variable to demonstrate the ability to alter the control flow and try during debugging alternative paths. In general, by modifying program variables and live replaying the affected control flows one can experiment during debugging with such features. In the simple program of Figure 28, the branch depends on the value of return_zero variable. As directly observed, the code follows the wrong control flow path (let us assume that the variable is named thoughtfully), since the program returns *1* when `return_zero` is true and vice versa. Clearly, the example is extremely simple, but the error is a typical mismatch between intended control-flow and the one eventually coded. Next we study the chances for experimenting with live corrections, during debugging.

*5.1.2.1 Changing with Forward Debugging*

After the forward debugger passes over the condition check, there is no way for the user to follow the correct code path and they would have to rerun the program up to the point before the condition, in order to be able to modify its value and divert the code towards the correct path and quickly test it. Certainly, due to triviality of the bug one need not rerun the scenario. But the general case of detecting the faulty branch logic is not that simple, with rerun commonly mandated. Alternatively, in this contrived case, manually moving the program counter (GDB and LLDB feature) to a previous address and modifying the value of the `return_zero` variable, would probably work sufficiently well, however that would not be possible in a slightly bigger and more complex than this trivial case program.

*5.1.2.2 Changing with Offline Reverse Debugging*

Offline reverse debuggers work on a prerecorded execution trace, thus, even though it is possible to step backwards, there is no way to modify the state of the program and replay following this time a different code path. Overall, state and control-flow are immutable in record-and-replay sessions.

*5.1.2.3 Changing with Live Reverse Debugging*

On the other hand, a live reverse debugger works on a live target and is able to step backwards, modify the state of the target and then continue forward, via a potentially different path. This scenario reflects the key synergy between reverse and forward debugging that, as we mentioned very early in Section 1.2.2 *Live Reverse Debugging, is one of the greatest advantages of live reverse debugging*. This is very briefly shown for our base scenario through the following actions:

Firstly, via typical forward debugging, we step forward just before the wrong `return` statement is reached, as shown in the image below.

```
* thread #1, queue = 'com.apple.main-thread', stop reason = step
    frame #0: 0x0000000100000fa5 branch`main at branch.cpp:7:9
   4        if (return_zero) {
   5            return 1;
   6        } else {
-> 7            return 0;
   8        }
   9    }
```

Then, by deploying the reverse debugger, we use the command `thread step-back-inst` to step-back two instructions (shortcut used as `pi -c 2` instead here), aiming to reach back the execution point just before testing of the `return_zero` variable is evaluated (i.e. beginning of branch test instruction). This is actually shown in the next image.

```
(lldb) pi -c 2
* thread #1, queue = 'com.apple.main-thread', stop reason = step
back 2 instructions
    frame #0: 0x0000000100000f8f branch`main at branch.cpp:4:9
   2          bool return_zero = false;
   3
-> 4          if (return_zero) {
   5              return 1;
   6          } else {
   7              return 0;
   8          }
```

Then, using the LLDB expression evaluation feature, we set `return_zero` to `true`.

```
(lldb) e return_zero = true
(bool) $0 = true

(lldb) s
```

Finally, we step forward as normal to observe what path is now taken once the value is changed to `true`. Since now it leads to `return 1`, though `return_zero` is `true`, it clearly indicates a flag-value / return-value mismatch, being adequate evidence to eventually resolve the error directly in code.

```
(lldb) s
Process 66961 stopped
* thread #1, queue = 'com.apple.main-thread', stop reason = step
in
    frame #0: 0x0000000100000f99 branch`main at branch.cpp:5:9
   2          bool return_zero = false;
   3
   4          if (return_zero) {
-> 5              return 1;
   6          } else {
   7              return 0;
   8          }
```

## 6. RELATED WORK
### 6.1 Software-Based Reverse Debuggers

Reverse debuggers based entirely in software make up the majority of current implementations and vary considerably in terms of their scope, as we are going to discuss below.

#### 6.1.1 Purely Functional Languages

Purely functional programming languages use persistent data structures, i.e., data structures that are never actually modified, but always copied on each write. Due to the nature of such programming languages, providing reverse debugging support requires less effort, since the debugger can simply make sure that all instances of a data structure remain alive during the duration of the debugging process, instead of tracking memory modifications. This similarly applies to other functional programming languages that are not identified as *pure*, but may provide analogous guarantees. A few notable implementations of reverse debuggers for functional languages are Elm Reactor [9] and the OCaml Debugger [12].

### 6.1.2 Unmanaged Runtimes

The term *unmanaged runtime environments* refers to programming language runtimes that support programs commonly consisting of native code and do not provide support for automatic memory management, e.g., garbage collection, or other features, such as thread scheduling.

Methods for recording and replaying or executing such target programs in reverse have been mentioned under Section 1.2.1*Offline Reverse Debugging*. Existing free and commercial reverse debuggers for unmanaged runtimes are the aforementioned *UndoDB* [6], the GNU Debugger (*GDB*) [13] and the Mozilla *rr* [14].

### 6.1.3 Managed Runtimes

On the opposite side, *managed runtime environments* are language runtimes responsible for interpreting source code or bytecode and generally provide at least automatic garbage collection. Historically, there have been multiple implementations of reverse debuggers for managed runtimes, adopting the record-and-replay approach. Since all of them are typical offline reverse debuggers, we enlist some representative systems according to the programming language runtime they are targeting:

- *JavaScript*: WebReplay [15] for WebKit, Replay [16] for Firefox, and the Firefox-based Web Replay [17] (being a different system to Replay)
- *Java*: Chronon [18]
- *Python*: RevPDB [19]

### 6.1.4 Virtual Machines

Reverse debuggers for (full system) virtual machines are able to track and back-track the whole operating system, thus enabling support for debugging OS kernels. Work on this area has been conducted, however, to our knowledge no reverse debuggers for virtual machines exists or is in use today. The most notable example has been the earlier support for record-and-replay for user-level programs in VMWare Workstation [20] in the period between 2008 and 2011.

### 6.1.5 Full-System Emulators

Full-system simulators are software systems that simulate both software and the underlying hardware, such as QEMU [21] or SPIM [22]. However, the most characteristic and notable full-system simulator, with live reverse debugging support, is Simics [24], which is able to undo and redo changes not only in memory but also in persistent storage and the CPU clock.

### 6.1.6 Program-Specific Solutions

All of the solutions mentioned up to this point were generic reverse debuggers that support different target programs. However, there have also been reverse debugging solutions implemented for particular software systems as a part of the software itself. One such approach is LLDB *Reproducers* [25], which are instrumentation statements injected at the entry points of the LLDB debugger's scripting API. Those statements record a trace containing the commands that were issued during a debugging session, along with their arguments and information about the debuggee. In case the debugger crashes or generates an internal error, the user can send the trace to the LLDB developers for investigation.

Another system is Whylines [23] for Java applications, which provides a *user interface* for user questions of the form *why this happened*. It relies on custom byte-code instrumentation to produce a *trace* holding all class files and execution history into a single data structure. Then trace is used to query, extract and chain selected calls, and providing informative views to the user. The tool blends instrumentation and recording to produce a structured named *trace*, although it is not a debugger system, neither provides reverse debugging features, being mainly an interactive history-analysis facility for forward debuggers.

*6.2 Hardware-Based reverse Debuggers*

Recording the execution of a program can also be facilitated or even completely offloaded to the underlying hardware, leading to much lower performance overhead.

*6.2.1   Processor Tracing Facilities*

Modern processor microarchitectures contain registers that can be leveraged for profiling or tracing, such as performance counters. The level of information and coverage of such registers varies a lot between microarchitectures and even processor families. An example technology is Intel Processor Trace [26], which captures a log of the instruction flow with a low overhead, allowing a debugger to later reconstruct the execution. This technology can be used to recover from a corrupted call stack, but does not trace modifications to registers and memory, so it not a complete reverse debugging solution, but rather an aid for reverse debuggers.

*6.2.2   Tracing via Specialized Hardware*

In a similar manner to full-system simulators, specialized hardware devices exist for recording not only the execution of a target program, but the whole system beneath it. These solutions are more common in the embedded and real-time system universe and a notable example is the Green Hills SuperTrace Probe [27].

*6.3 Hybrid Approaches*

Besides software- or hardware-only solutions, recent advancements in reverse debuggers have combined processor tracing facilities with software-generated core dumps, in order to reconstruct the execution of the target program. One such approach has been followed in REPT [28] by Microsoft Research, using an execution trace from Intel Processor Trace ([26], mentioned earlier in Section 6.2.1 *Processor Tracing Facilities*), and using WinDbg as its frontend, with trace information collected and reported using Windows Error Reporting (WER) [10].

7. SUMMARY, CONCLUSIONS AND FUTURE WORK
We have outlined the full-scale implementation of live reverse debugging features on top of LLDB, relying on the runtime capturing of state snapshots, by working intensely, with the inherent performance implications, at single-step CPU mode.

Based on this, we implemented the reverse counterpart of a wide range of typical tracing debugging operations, while setting the reverse extension to work in collaboration, and effectively on top of, the primary forward LLDB debugger.

We consider that extra future work is necessary to bring it to production level that LLDB itself is, along the following lines at least: (a) minimizing the memory footprint of the taken snapshots, for instance, by only copying blocks that include modified content (but with the expense of maintaining linked data structures); (b) adding support for watchpoints triggered when executing in reverse mode; (c) offering more operations and relevant commands for the developers to inspect the contents of recorded tracepoint snapshots; (d) expose the reverse debugging API through the public API of LLDB, in order to enable other programs, especially IDEs, to take direct advantage of the functionality in the debugging User-Interface; and (e) to provide support for multi-threaded programs, something that we consider is a big chapter by its own, and we did not investigate at all as part of this work.

When we firstly started this development effort there was an initial big concern as to the utility of live reverse debugging in real-life situations. The complexity of present systems and the fact that they use multiple resources, and thus generating thousands of non-deterministic events in a short operation of time (effectively, those non-reversible), was an important discouraging factor. But then, once we decided to focus on the actual debugging practices, it became apparent that the prevalent debugging style emphasizes the *detection of the smallest possible offensive case scenario, so that an ideally minimal system need only be traced and examined*. This principle essentially reflects the overall art of debugging, meaning that just a subset of a subset of a real system is effectively put under the analytic lenses of a debugger.

Although this does not solve the non-deterministic events issue, or the problem with multi-threaded systems, it helped us take two important decisions. Firstly, to *make the operation of the live reverse debugger a special mode of a forward debugger that can be switched on and off on-demand*. Secondly, to defer performance and memory improvements at a second development phase, putting *primary focus on the modular integration with an existing forward debugger and the delivery of wide set of reverse tracing operations*.

Overall, reverse debuggers are very complex tools to develop, being still the subject of many on-going research efforts, mainly led by industry. In particular, live reverse debugging represents a newest, less-explored category that procedurally blends very well with forward debugging, offering an added-value operational mode. Clearly, live reverse debugging is not meant or suggested to eventually substitute record-and-replay debuggers. However, it can help a lot by enabling more exploratory debugging practices, when things get developed and tested first-time in isolation, as it is the case with unit testing where everything is worked and polished at a smaller scale, in isolation, with *unit debugging*. Thus, we consider there is a lot of fertile ground for reverse debuggers to further grow and mature, and believe our work is a milestone in this direction, and hopefully can provide useful implementation material for the developers of the next generation of far better live reverse debuggers.

## 8. CONFLICTS OF INTEREST
Authors have no conflict of interest relevant to this article.